\documentclass[12pt,3p]{elsarticle}

\usepackage{multicol}
\usepackage{color}
\usepackage{xspace}
\usepackage{enumerate}
\usepackage{lineno}
\usepackage{graphicx}
\usepackage{float}
\usepackage{amsmath}
\usepackage{amssymb}

\begin{document}
\begin{frontmatter}

\title{A Method to Load Tellurium in Liquid Scintillator for the Study of Neutrinoless Double Beta Decay \\
{\small \it (NIM A,Volume 1051, 2023, 168204 )}}

\author[label1]{D.J.~Auty}
\address[label1]{University of Alberta, Department of Physics,
4-181 CCIS,  Edmonton, AB T6G 2E1, Canada}

\author[label6]{D.~Bartlett}
\address[label6]{Queen’s University, Department of Physics, Engineering Physics \& Astronomy, 64 Bader Lane, Kingston, ON K7L 3N6, Canada}

\author[label2]{S.D.~Biller \corref{cor1}}
\address[label2]{University of Oxford, Department of Physics,
Denys Wilkinson Building,  Keble Rd, Oxford OX1 3RH, UK}
\cortext[cor1]{corresponding author}
\ead{steven.biller@physics.ox.ac.uk}

\author[label5,label6]{D.~Chauhan}

\author[label6]{M.~Chen}

\author[label3]{O.~Chkvorets}
\address[label3]{Laurentian University, Department of Physics,
935 Ramsey Lake Road, Sudbury, ON P3E 2C6, Canada}

\author[label6]{S.~Connolly}

\author[label6]{X.~Dai}

\author[label6]{E.~Fletcher}

\author[label4]{K.~Frankiewicz}
\address[label4]{Boston University, Department of Physics,
590 Commonwealth Avenue, Boston, MA 02215, USA}

\author[label4]{D.~Gooding}

\author[label4]{C.~Grant}

\author[label5]{S.~Hall}
\address[label5]{SNOLAB, Creighton Mine No. 9,
1039 Regional Road 24, Sudbury, ON P3Y 1N2, Canada}

\author[label6]{D.~Horne}

\author[label9]{S.~Hans}

\author[label6]{B.~Hreljac}

\author[label7,label8]{T.~Kaptanoglu}
\address[label7]{University of Pennsylvania, Department of Physics \& Astronomy, 209 South 33rd Street, Philadelphia, PA 19104-6396, USA}
\address[label8]{University of California, Berkeley, Department of Physics, 366 Physics North MC 7300, Berkeley, CA 94720-7300, USA}

\author[label6]{B.~Krar}

\author[label3,label5]{C.~Kraus}

\author[label2,label7]{T.~Kroupov\'a}

\author[label6]{I.~Lam}

\author[label6]{Y.~Liu}

\author[label9]{S.~Maguire}
\address[label9]{Brookhaven National Laboratory, Chemistry Department, Building 555, P.O. Box 5000, Upton, NY 11973-500, USA}

\author[label6]{C.~Miller}

\author[label5,label6]{S.~Manecki}

\author[label9]{R.~Rosero}

\author[label2]{L.~Segui}

\author[label1]{M.K.~Sharma}

\author[label5]{S.~Tacchino}

\author[label6]{B.~Tam}

\author[label6]{L.~Tian}

\author[label1]{J.G.C.~Veinot}

\author[label3]{S.C.~Walton}

\author[label10]{J.J.~Weigand}
\address[label10]{Technische Universität Dresden, Faculty of Chemistry and Food Chemistry, 01069 Dresden, Germany}

\author[label6,label11]{A.~Wright}
\address[label11]{Institute of Particle Physics, Canada}

\author[label9]{M.~Yeh}

\author[label6]{T.~Zhao}

\begin{abstract}
A  method has been developed to load tellurium into liquid scintillator 
so as to permit searches for neutrinoless double beta decay with high sensitivity. The approach involves the synthesis of an oil-soluble tellurium compound from telluric acid and an organic diol. The process utilises distillable chemicals that can be safely handled underground and affords low radioactive backgrounds, low optical absorption and high light yields at loading levels of at least several percent Te by weight. 
\end{abstract}


\end{frontmatter}

\section{Introduction}
The search for neutrinoless double beta decay ($0\nu\beta\beta$) is recognized as one of the highest priority areas of modern particle physics research. The observation of this lepton-number violating process would establish the absolute mass scale of neutrinos, provide insight into Grand Unification and lend weight to models of leptogenesis. \cite{theory}. 

One promising technique being employed by the SNO+ experiment \cite{SNO+} 
for such a search involves loading a large volume of liquid scintillator based on linear alkylbenzene (LAB) with a candidate $0 \nu \beta \beta$ isotope. The use of large self-shielding volumes of high purity liquids allows the potential for realising very large quantities of relevant isotope with low levels of background. Further, the ability to easily modify the liquid configuration and separately assay different components offers a significant advantage for probing any candidate signal and readily allows the implementation of further improvements to purification and loading. A particularly interesting candidate isotope in this regard is $^{130}$Te. Owing to the isotope's large natural abundance (34\%), enrichment is unnecessary, making it considerably more economical and practical to deploy large quantities.

A method for loading tellurium into organic liquid scintillator has been developed based on the formation of oil-soluble 
compounds derived from telluric acid (Te(OH)$_6$, hereafter TeA) and organic diols in conjunction with N,N-dimethyldodecylamine (DDA), which acts as a stabilisation/solubilisation agent. While the method is amenable to a variety of diols, 1,2-butanediol (BD) was selected due to its relatively low molecular weight (permitting a higher fractional Te content), relatively low cost and ease of distillation. The chemicals involved can thus all be purified to high levels, have high flash points and are relatively safe to work with in underground environments. In addition, both BD and DDA can be obtained via non-biogenic production processes, which is important in order to maintain the low $^{14}$C levels needed to allow low threshold operation of large liquid scintillation detectors. The loading process results in low optical absorbance and good light output in larger detectors for loading levels of at least several percent Te by weight. This is particularly important for $0 \nu \beta \beta$ experiments, where good energy resolution is needed to reduce backgrounds from $2 \nu \beta \beta$ events. Stability of the loading has been explicitly demonstrated to be at least in excess of 5.5 years at room temperature. In addition, acrylic compatibility of the loading technique was established via accelerated ageing tests, in which stressed acrylic pieces were exposed to highly concentrated ($\sim$25\%) Te-loaded LAB. No additional crazing or deterioration of the mechanical integrity was observed within the sensitivity of the tests. Two variants of the loading method are presented below, each resulting in slightly different LAB-soluble mixtures of tellurium compounds with different fluorescence quenching properties. Sources for the TeA\cite{TeA}, BD\cite{BD} and DDA\cite{DDA} used for these studies are provided in the references.

\section{Loading Method}

\subsection{Overview}
	
When BD is added to dilute aqueous solutions of TeA at room temperature, condensation reactions lead to the attachment of up to two bidentate BD ligands for each tellurium center, as indicated by nuclear magnetic resonance (NMR) measurements in Figure 1. This dilute form with 2 attachments will hereafter be referred to as DF2.

\begin{figure}[H]
\centering
\includegraphics[width=140mm]{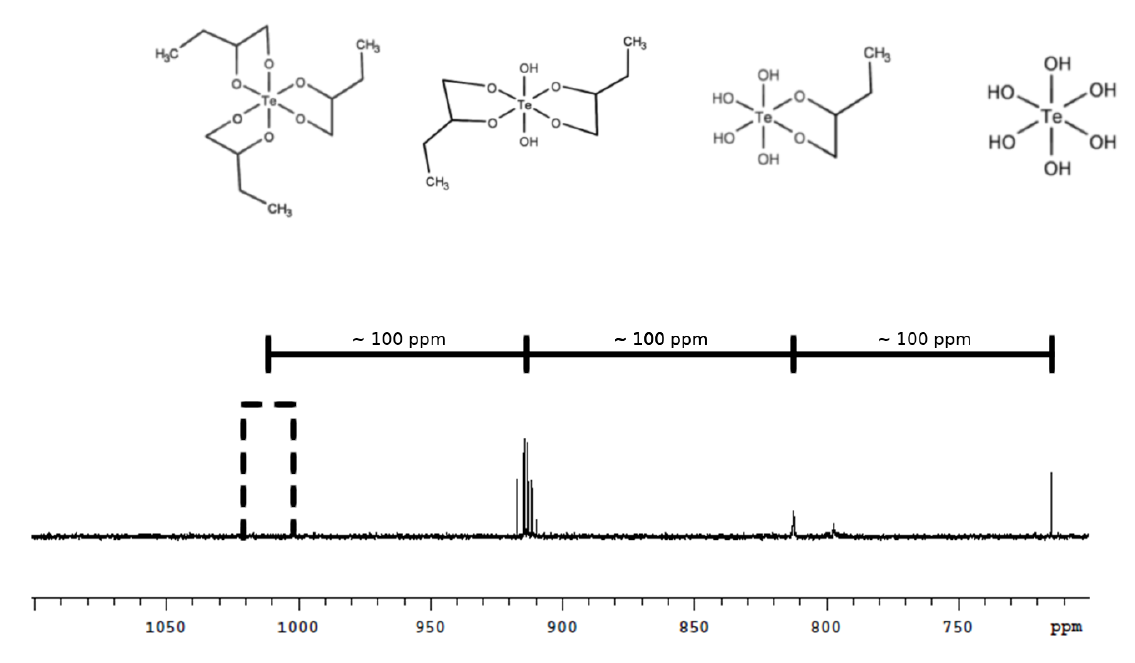}
\caption[]{NMR spectra showing spontaneous formation of ``precursor'' TeA-diol forms for dilute concentrations of TeA and BD in water at room temperature. Representative forms are shown at the top. No peaks corresponding to 3 attached diols are seen (dashed box), implying a preference for 2 bidentate attachments.}
\label{DiluteForms}
\end{figure}

This condensation process is in equilibrium with hydrolysis reactions, as indicated in Figure 2.

\begin{figure}[H]
\centering
\includegraphics[width=150mm]{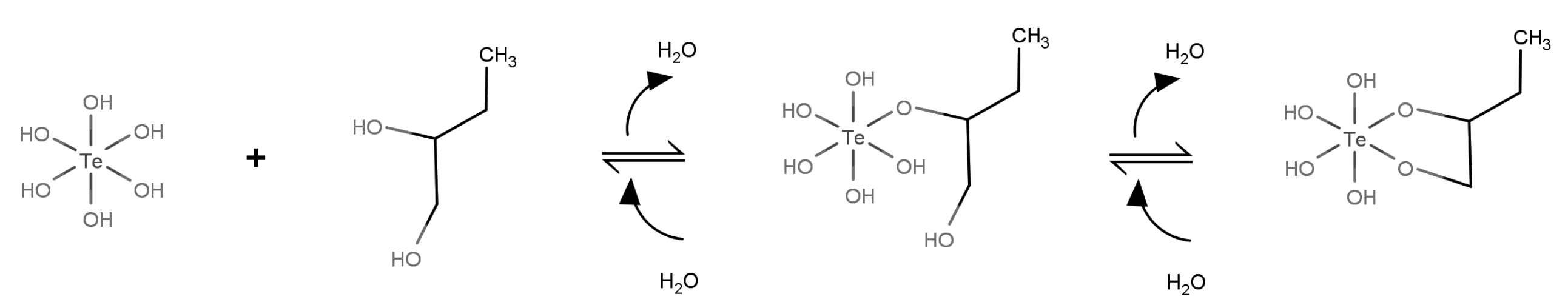}
\caption[]{Equilibrium between condensation and hydrolysis reactions for BD and TeA in water.}
\label{Equilibrium}
\end{figure}

At higher concentrations of TeA, the environment is more acidic; the TeA has a pH of $\sim$3.5 for a 10\% aqueous solution and the BD ligand attachment further increases the acidity. This tends to drive the equilibrium in the direction of hydrolysis. The balance can be tipped back towards condensation by adding N,N-dimethyldodecylamine (DDA) or heat to the system. Either approach thus leads to the formation of precursor compounds that form the basis for diol-loading of tellurium. Solubilisation in LAB can then be accomplished by either 1) adding additional heat to form more complex oligomers through further condensation reactions or 2) using an amine such as DDA for neutralisation in water and to then associate a long carbon chain through non-binding ionic interactions with the precursor compounds for solubility in LAB. Details of these two approaches and properties of the resulting Te- loaded scintillator (TeLS) mixtures will now be described. The following nomenclature will be used: PC1 will refer to the precursor compounds initially formed from the addition of heat; TeBD1 will refer to the oligomers formed from further heating of PC1; PC2 will refer to the precursor compounds initially formed via neutralisation; and TeBD2 will refer to the solubilised pairing of PC2 and DDA.

\subsection{Type I Loading: Solubilisation via Heating}

In this approach, aqueous TeA (30-50\% w/w) is combined in a 1:3 molar ratio with BD. $^{125}$Te NMR and mass spectrometry measurements suggest that the initial PC1 formations are largely  Te monomers with two bidentate and one monodentate BD attachments.  The mixture is heated under vacuum to remove water at 70-80$^{\rm o}$C, with care taken to avoid local temperatures above 90$^{\rm o}$C to inhibit side reactions. The mixture is agitated and sparged with nitrogen gas to promote water removal. As water is removed from the reaction system, additional condensation reactions occur. This leads to the formation of a viscous and complex mixture of short-chain oligomers in which the precursor compounds are more fully substituted and linked together by BD. This is evident in the $^{125}$Te NMR spectra as broadening of peaks, commonly seen in polymerisation/oligomerisation, which re-form singlet peaks when the proton is decoupled (Figure 3). Extensive oligomer formation is not observed if water is not removed from the reaction, which is consistent with an equilibrium between hydrolysis and condensation reactions. With continued heating and water evaporation, the viscosity of the product increases. This observation is suggestive of increasing molecular weight via oligomerisation and/or self-interaction of the oligomers.

As the heating and water removal continues, the TeBD1 product becomes soluble in LAB. The transition to solubility is rapid; the material changes from largely insoluble in LAB to completely soluble in a very short period of time. Typical water concentration is extremely low at the point of solubility ($<$0.1\%). The sudden transition to solubility may be the result of structural self-organisation of the oligomer chains. At the LAB solubility point, the addition of a small amount of additional BD to the TeLS induces the formation of two phases. Continued heating of the LAB soluble TeBD material results in further evolution of the $^{125}$Te NMR (Figure 3), and increased ``BD tolerance''. Tolerance of BD addition can thus provide a useful diagnostic of reaction progress. This is likely the result of structural evolution involving continued dehydration within the oligomers. 

\begin{figure}[H]
\centering
\includegraphics[width=120mm]{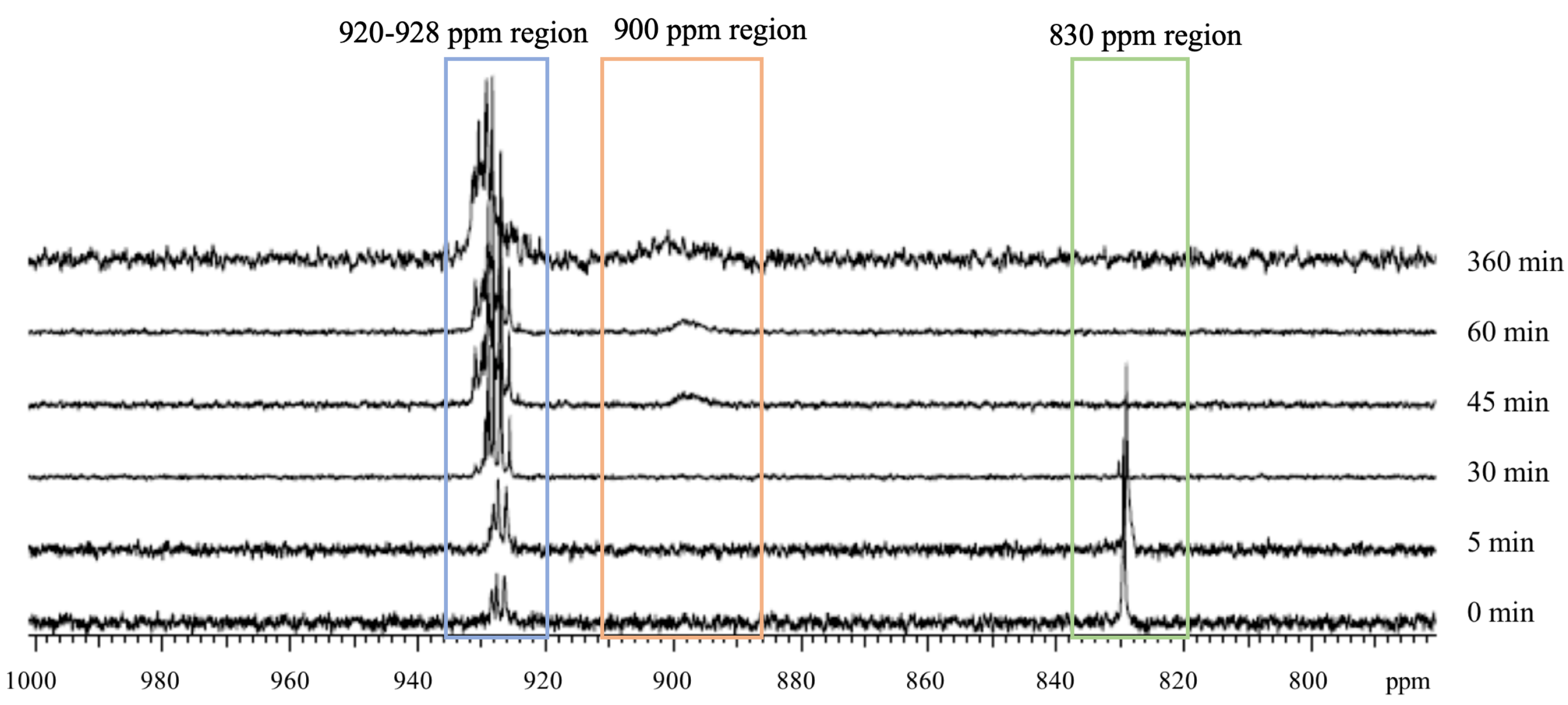}
\caption[]{$^{125}$Te NMR spectra of TeBD1 formed by oligomerisation as a function of heating time.}
\label{SOP_NMR}
\end{figure}

Electrospray ionisation (ESI) Mass Spectrometry (MS) data for TeBD1 produced with a 3:1 BD:TeA ratio is consistent with multiple species consisting of short oligomer chains containing up to at least three Te centers (Figure 4). 

\begin{figure}[H]
\centering
\includegraphics[width=120mm]{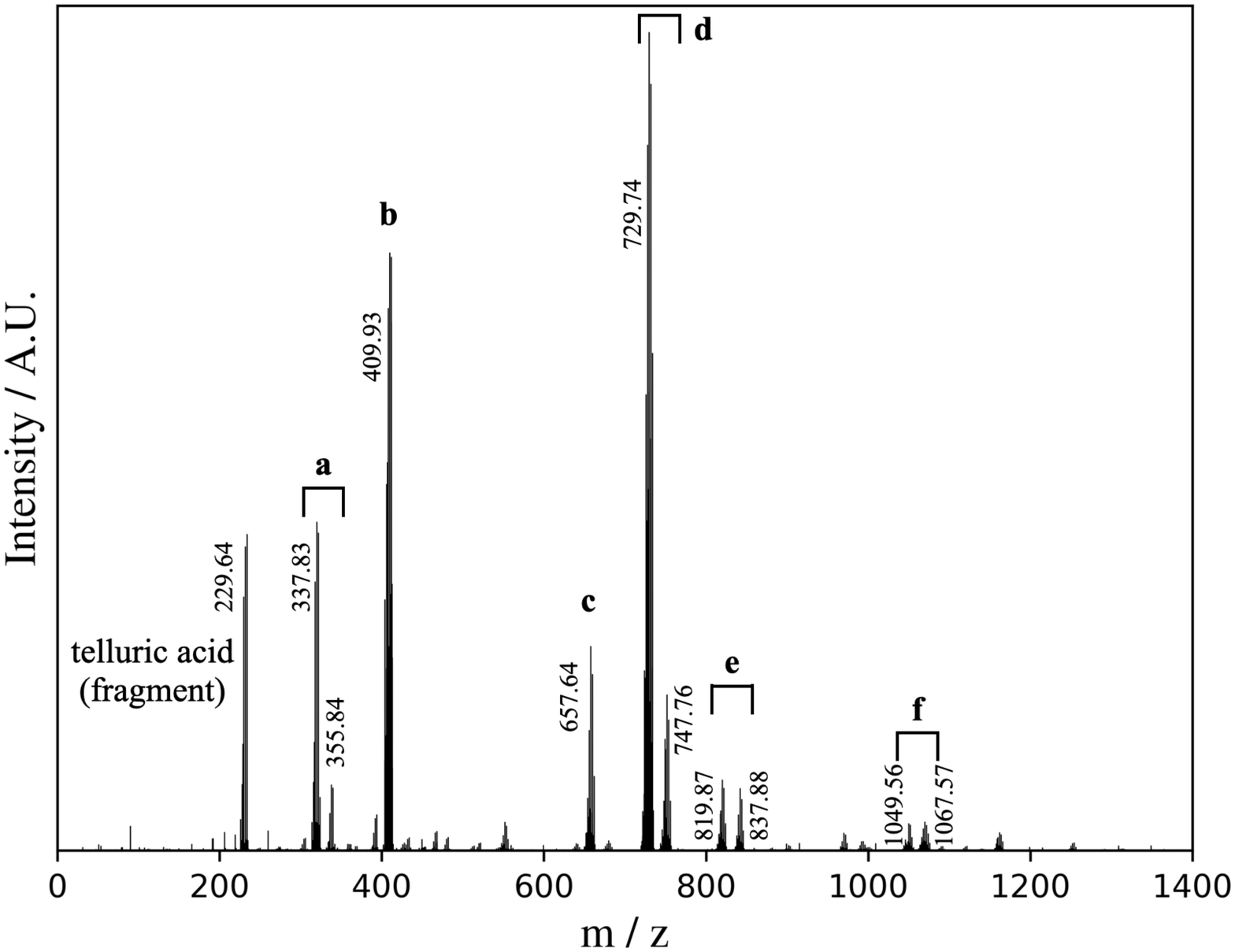}
\caption[]{Positive ion ESI mass spectra for TeBD1 synthesis via oligomerisation at 80$^o$C. The sample was heated for two hours beyond the LAB solubility point. Possible structures for various peaks are shown in Figure 5.}
\label{ESI}
\end{figure}

Using the observed masses and elemental distributions, it is possible to identify potential configurations of the various species in the MS spectra. Illustrations of several such structures corresponding to the main peaks in Figure 4 are shown in Figure 5.

\begin{figure}[H]
\centering
\includegraphics[width=150mm]{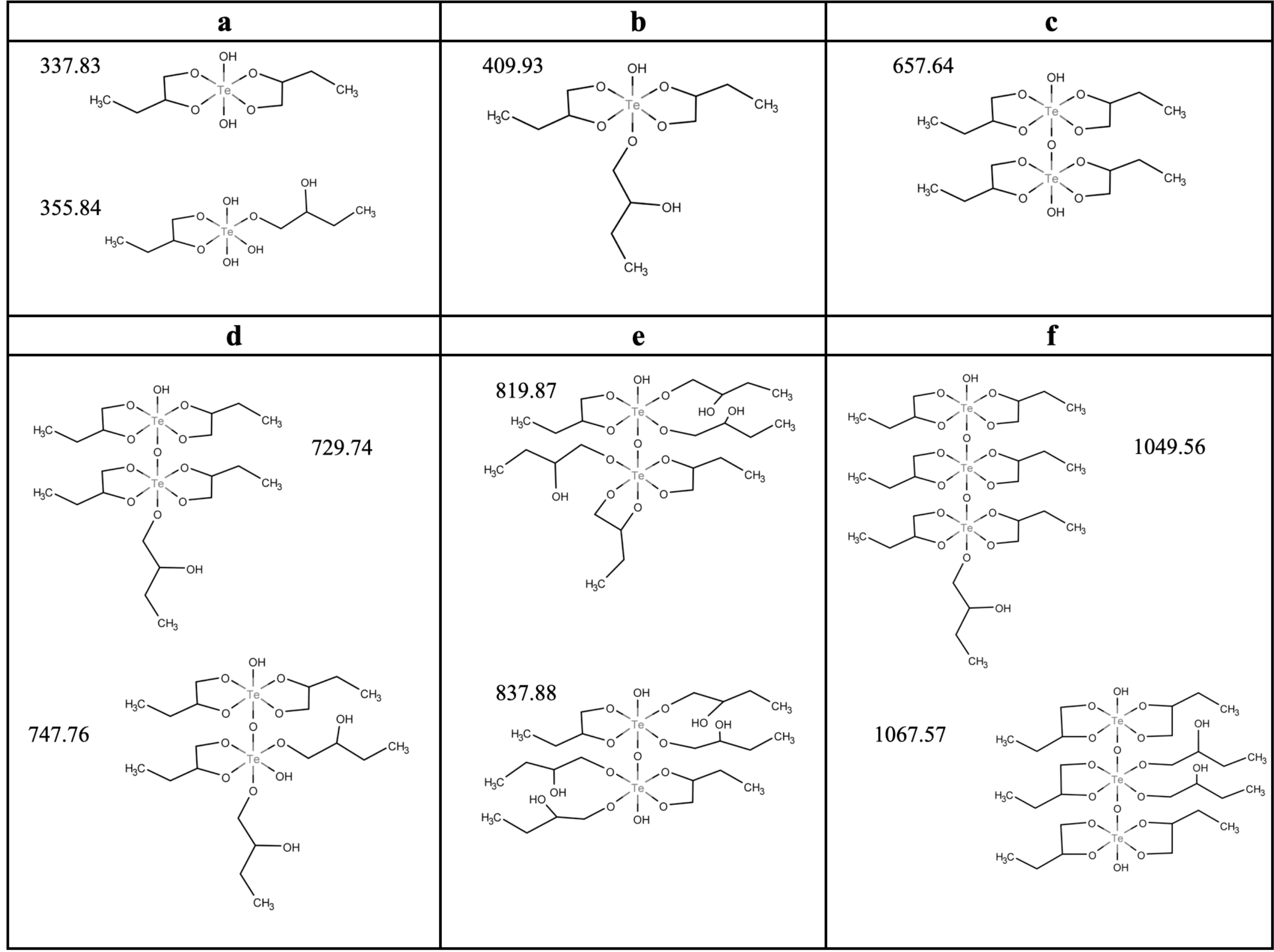}
\caption[]{Possible structures of dominant species identified in Figure 4.}
\label{Oligomers}
\end{figure}

The integrity of the TeBD1-LAB mixture is highly sensitive to the presence of water. Indeed, over the long term, even tiny amounts of water ($\sim$hundreds of ppm) can lead to the deterioration of TeBD1 oligomers as the result of hydrolysis reactions that eventually lead to a loading instability, where white crystalline Te-containing material appears either in the liquid itself or attached to the glass of the sample container (see Section 3). In a carefully controlled environment, TeBD1 in LAB can be stable for several years or more. Nevertheless, it is advantageous to add a stabiliser to prevent crystal formation and provide an environmentally robust loading protocol. 

\subsection{Solubilisation and Stabilisation with DDA}

The stability of TeBD1 in scintillator can be dramatically improved by the introduction of DDA.  The principle solubilisation mechanism is believed to be the formation of an association between OH groups in the Te compounds and tertiary amino groups of DDA, effectively adding a long hydrocarbon tail to the compounds, thereby promoting solubility in LAB and inhibiting further oligomerisation. Tests indicate that DDA-protected TeLS is very stable and robust against water exposure, as detailed in Section 4. The de-protonation of the acidic oligomers by the amine has the added benefit of reducing the fluorescence quenching in the scintillator. Based on observations of the light yield, the optimal molar ratio of DDA to Te appears to be close to 0.5:1, with a fairly broad maximum extending from $\sim$0.25-0.75 (Figure 6). Ratios in this range were also found to work well for maintaining stability and solubilising crystals in samples that had been exposed to water prior to DDA addition.

\begin{figure}[H]
\centering
\includegraphics[width=120mm]{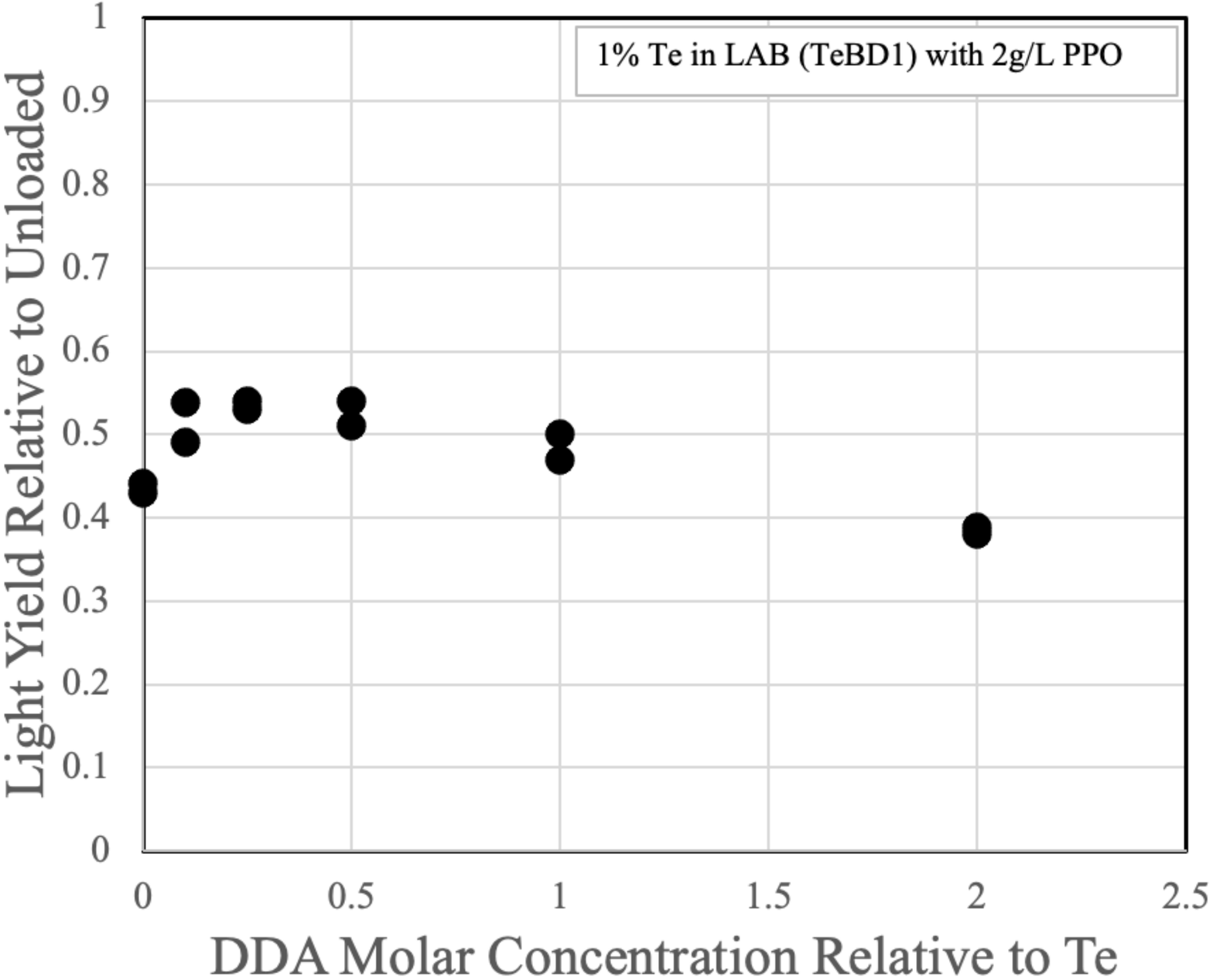}
\caption[]{Relative light level as a function of DDA molar concentration with respect to Te for 1\% loading in LAB with 2g/L PPO.}
\label{DDA_Max}
\end{figure}

\subsection{Type II Loading: Direct Solubilisation of PC2 Formations with DDA}

Neutralisation and solubilisation of the precursor compounds in LAB can be accomplished by first adding DDA to the initial reaction mixture with a molar ratio of 0.5:1:2 for DDA:TeA:BD. In this case, formation of the precursor compound (PC2) happens even in the absence of heat and that compound is then directly solubilised into LAB via DDA through the ionic association described above without the need for oligomerisation. This ``cold synthesis'' approach, which need not take place in an aqueous environment, simply requires the removal of water generated by the condensation reactions that take place during compound formation in order to mix with LAB. This can be accomplished, for example, via nitrogen sparging. Mass specrometry and NMR measurements confirm that, when no heat is applied, the  PC2 compounds that are solubilised to form TeBD2 consist almost entirely of tellurium monomers with two bidentate BD attachments (Figures 7-8). These are identical to DF2 formations in Figure 1, indicating a relatively weak association between almost all tellurium species and DDA.

\begin{figure}[H]
\centering
\includegraphics[width=140mm]{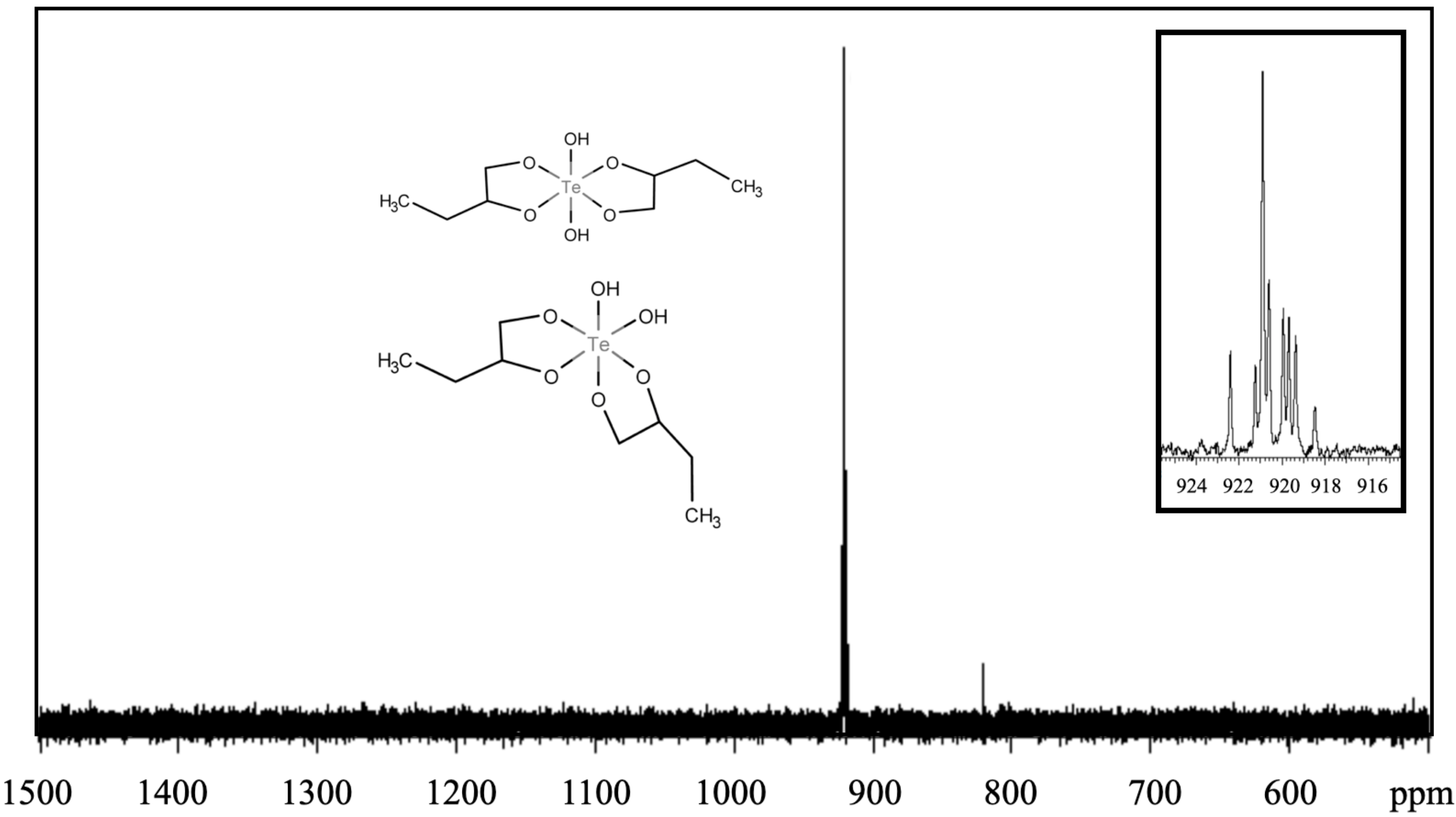}
\caption[]{NMR spectrum of TeBD2 formed via direct DDA solubilisation with a 2:1 BD:Te ratio. Inset shows a zoomed view of main peak.}
\label{NMR_DDA1}
\end{figure}

\begin{figure}[H]
\centering
\includegraphics[width=140mm]{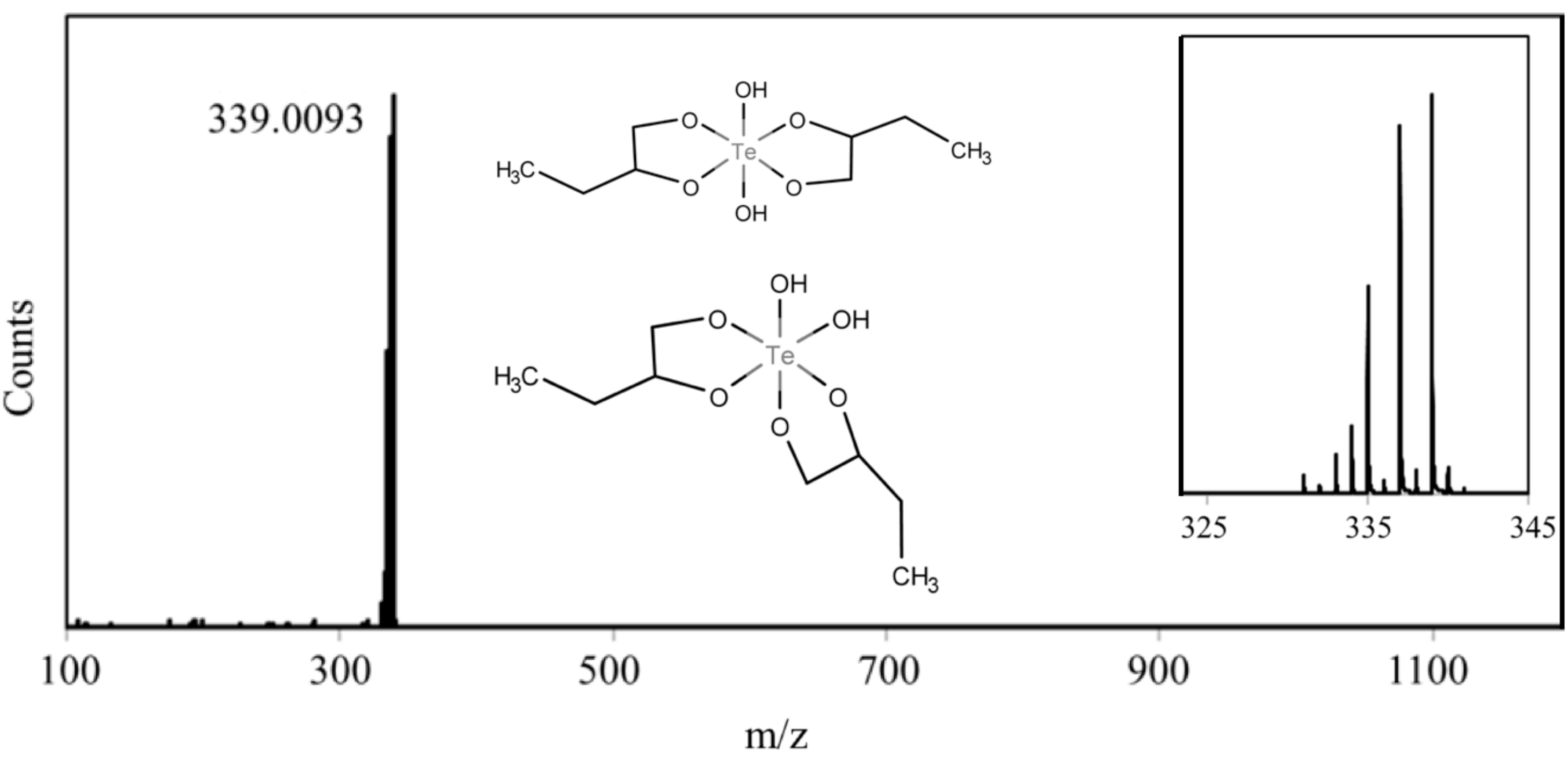}
\caption[]{ESI mass spectrum of TeBD2 formed via direct DDA solubilisation with a 2:1 BD:Te ratio. Inset shows a zoomed view of main peak.}
\label{Mass_DDA}
\end{figure}

It is of particular interest to note that the fluorescence quenching in this approach is significantly reduced relative to oligomerisation, as discussed in Section 4. This would allow for a notable increase in loading levels while maintaining the same light output (and, hence, energy resolution). 

\subsection{Hybrid Loading}
We briefly note in passing that mixed loading approaches are possible, where heating is initially used to begin precursor formation and drive off the majority of the water, at which point DDA is then employed to complete the solubilisation. This is the approach being taken by SNO+, as it allows TeA to be transported as a solution during the purification and synthesis, while also permitting a more controlled and rapid solubilisation with DDA. The nature of the forms produced and their fluorescence quenching characteristics (described below) are, perhaps unsurprisingly, between those of Type I and Type II approaches, with the specific mix dependent on the amount of heating used before DDA is introduced.

\section{Loading Stability}

Over the course of TeLS development, many hundreds of samples have been produced under various conditions and monitored for stability over several years. The loading has been found to be robust with regard to environmental conditions, aside from the addition of large amounts of acid or base or exposure to water. As previously mentioned, TeBD1 that has not been stabilised with DDA is highly water sensitive and even modest humidity exposure ($\sim$hundreds of ppm) can result in crystal formation on the timescale of weeks to months, depending on the degree of exposure. X-ray crystallography indicates these formations are composed of DF2 units. However, none of the samples containing DDA have ever shown signs of such behaviour after more than 5.5 years of monitoring.

Separation effects have also been observed following humidity exposure, particularly in samples without DDA, in which a heavier gel-like portion containing Te settles to the bottom of the sample vial. This is believed to be due to the association of water molecules with TeBD1 or TeBD2 rather than the immediate hydrolysis of the compounds themselves. This phenomenon can be sensitively tracked by monitoring the fluorescence yield of the mixture, which increases as phase separation progresses owing to reduced fluorescence quenching in the remaining portion of the scintillator. DDA dramatically reduces the propensity for phase separation, but does not eliminate it altogether in cases of extreme humidity exposure. However, even when separation in samples involving DDA does occur, loading integrity can be restored through the removal of water.

Studies of these separation effects were performed by standing open vials of loaded scintillator with 2g/L of 2,5-Diphenyloxazole (PPO) in a sealed container holding a $\sim$1~cm deep pool of water to create a high humidity environment. Water beads in the samples were visible after a day, indicating that they were saturated. The fluorescence yield was monitored by periodically exposing samples to a $^{90}$Sr$/^{90}$Y source and measuring the light output with a standard bialkalai photomultiplier tube. Figure 9 compares the endpoint of the observed fluorescence spectra after 300s of counting relative to the initial values as a function of time for 1\% Te loading with TeBD1 following a 48hr humidity exposure for samples with and without DDA.

\begin{figure}[H]
\centering
\includegraphics[width=130mm]{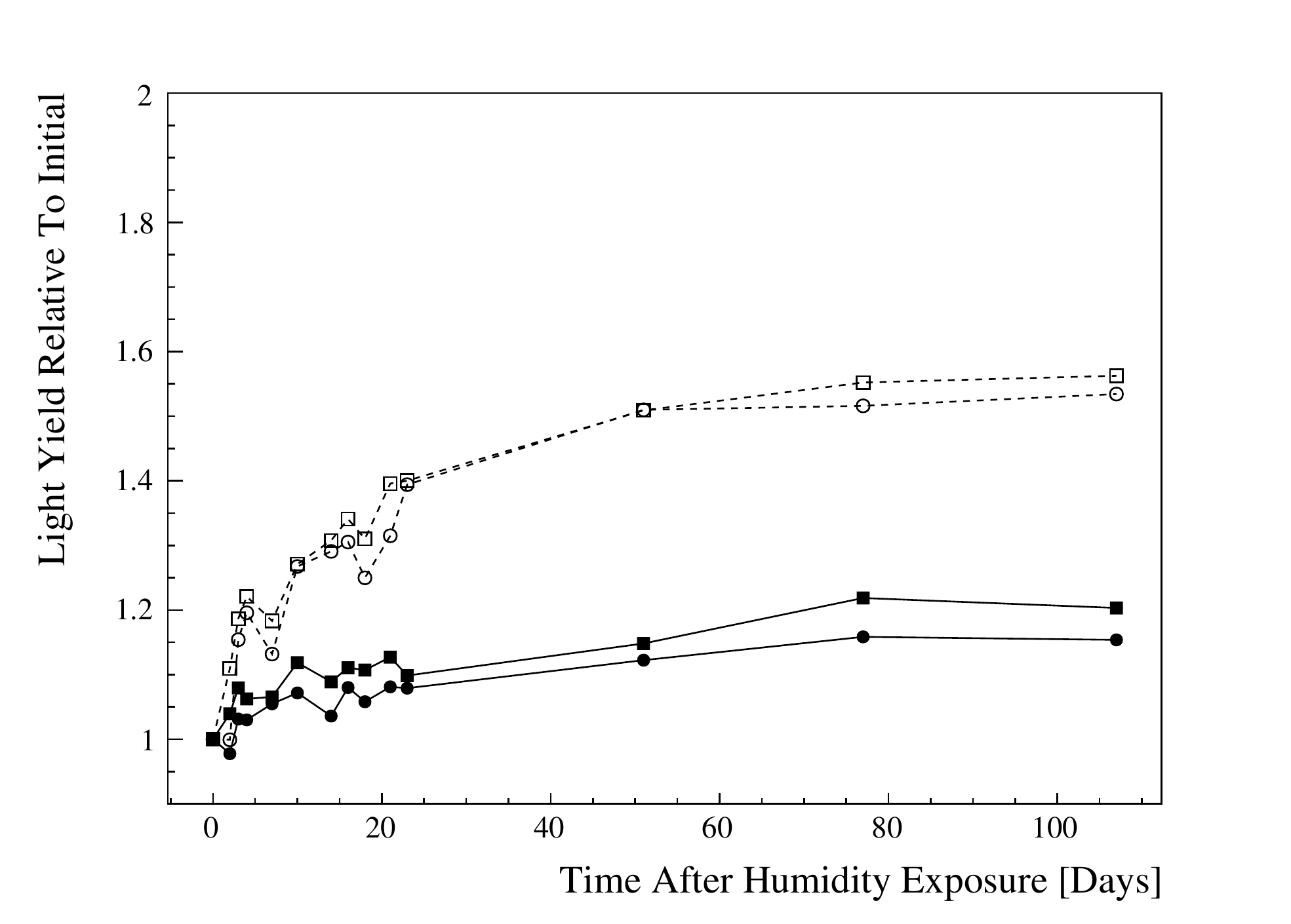}
\caption[]{Relative fluorescence levels (inversely related to Te content) for scintillator initially loaded with TeBD1 in a 1\% Te concentration as a function of time following a 48hr humidity exposure. Clear symbols represent samples without added DDA and solid symbols represent samples with DDA added. Data from duplicate samples of each type are shown to indicate the size of sample variations.}
\label{HumidityEndpoint}
\end{figure}

The behaviours of TeBD1 + DDA and TeBD2 were studied for a much more extreme exposure of 3 months. In both cases, the fluorescence yield after humidity exposure approaches that of unloaded scintillator. XRF measurements confirm that the majority of Te had been removed from the LAB phase at this stage. The samples were then sparged with nitrogen gas for two hours to remove the water. In both cases, XRF measurements indicated that full Te loading had been recovered. Interestingly, the post-recovery TeBD1 + DDA fluorescence spectrum exhibited a higher light yield compared to that before humidity exposure, but matched well with the spectrum from TeBD2 (which has an intrinsically higher light yield). This suggests that the larger TeBD1 formations may have been broken down by hydrolysis into the smaller structures that were then solubilised by DDA, effectively creating a TeBD2 loading.

\section{Optical Properties}

\subsection{Light Yield}
Figure 10 shows the relative light levels observed as a function of Te loading. This is based on inferring the endpoint of a $^{90}$Sr/$^{90}$Y source that is placed beneath a standard borosilicate fluorescence vial containing the scintillator sample and measuring the pulse height distribution observed using a  bialkalai PMT (Hamamatsu H11432-100). Measurement uncertainties are dominated by systematics associated with sample preparations and variations in the vial thickness, which alters the average energy of electrons that enter the scintillator. The results of multiple samples in different vials are therefore shown to indicate the spread in measurement values.  The trends are well described by the linear Stern-Volmer relationship, as shown in Figure 11. For LAB with 2g/L PPO, it is seen that fluorescence quenching to half the unloaded light level occurs at about 0.7\% Te for TeBD1 and $\sim$1.8\% Te for TeBD2. These figures also indicate that, by tripling the PPO concentration to 6g/L, this can be further improved, with light levels halving at concentrations of $\sim$3\% Te, representing an effective reduction in quenching by a factor of $\sim$3.5 relative to TeBD1 with 2g/L PPO.

\begin{figure}[H]
\centering
\includegraphics[width=140mm]{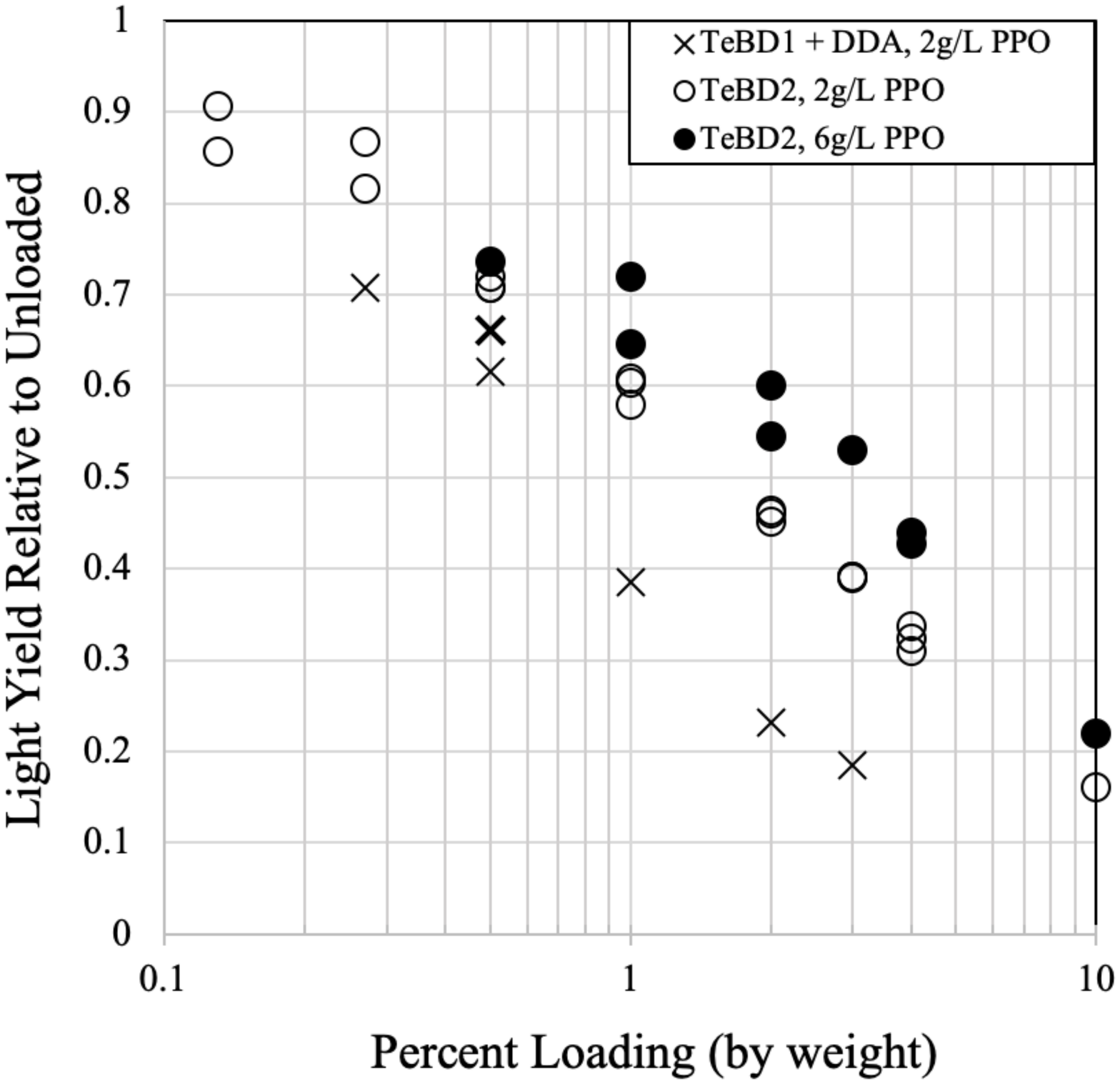}
\caption[]{Relative light yields as a function of percent Te loading for LAB with 2g/L PPO and TeBD1 loading (crosses); TeBD2 loading (open circles); and TeBD2 loading with boosted PPO (solid circles).}
\label{LightLevels}
\end{figure}

\begin{figure}[H]
\centering
\includegraphics[width=140mm]{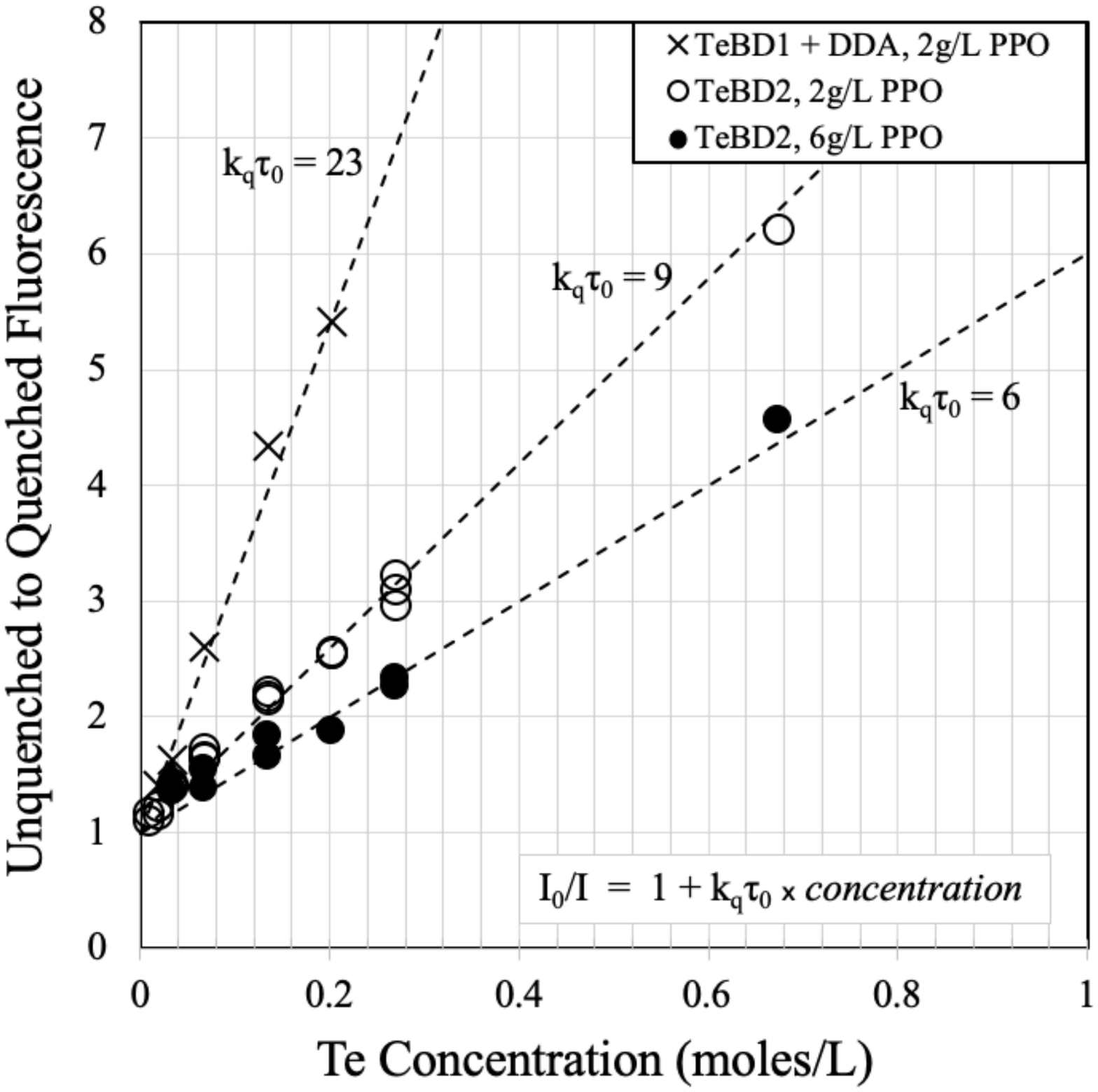}
\caption[]{Ratio of unquenched to quenched fluorescence yields as a function of Te concentration for LAB with 2g/L PPO and TeBD1 loading (crosses); TeBD2 loading (open circles); and TeBD2 loading with boosted PPO (solid circles).}
\label{QuenchingCurve}
\end{figure}

\subsection{Transparency}

The UV-vis extinction of TeBD compounds was measured using a variety of loaded LAB samples containing 10\% concentrations of Te. Some samples involved purely heated solubilisation and some used a mixture of heated and DDA solubilisation (no dependency on this was observed). In all cases, telluric acid was first purified through thermal recrystallisation and solutions were then filtered to remove insoluble impurities. The LAB, BD and DDA were also distilled. Extinction curves were obtained using a 10~cm quartz cell in a PerkinElmer Lambda-800 Spectrophotometer. These were then normalised in the region of 700~nm, far from the expected influence of TeBD absorption and where LAB transparency is high, so that the contribution of LAB could be accurately subtracted. Results were also cross-checked using a 5~cm cell. The resulting LAB-subtracted curves are shown in Figure 12.  These measurements are dominated by systematic uncertainties, with the observed variations between samples due to differences in sample preparation, including purification and filtering. Extinction lengths at higher wavelengths are also at the limit of our systematic uncertainties in subtracting the spectra. Nonetheless, excellent transparency is indicated, even at these high loading levels, with extinction lengths exceeding 10~m above $\sim$400~nm. The transparency at percent-level loadings is therefore expected to be significantly better than this.

A large number of samples (mostly with 0.5\% Te in LAB) have been monitored for long periods of time to establish long term optical stability. In a subset of cases, increased absorption was observed at lower wavelengths in the region below $\sim$430~nm. In addition to the known UV sensitivity of DDA, a dependence on the quality/age of the LAB and DDA was observed that related to the amount of oxidation and/or free radicals present at the time of sample preparation. If this activity is too high, this can lead to the observed absorption increase. However, for samples produced from freshly purified stock materials that were stored under dark conditions, no noticeable change in absorption was observed over a timescale of more than 3 years at room temperature. There is some indication that TeBD itself may act as an anti-oxidant to maintain optical stability if starting from initially fresh stock materials. The early introduction of antioxidants may also help further mitigate these effects (currently under study).

\begin{figure}[H]
\centering
\includegraphics[width=120mm]{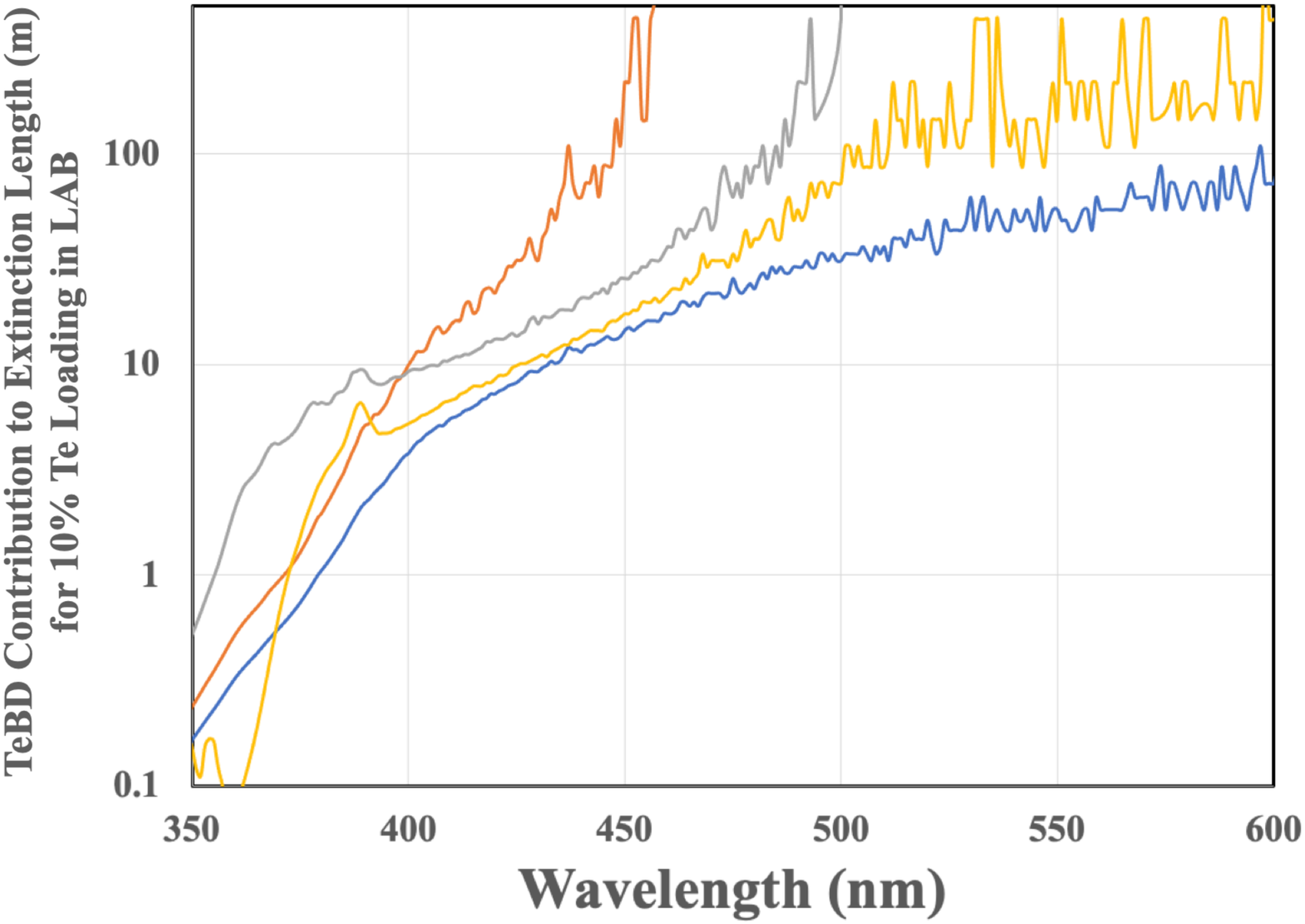}
\caption[]{Extinction length of TeBD for various 10\% Te samples with contributions from LAB subtracted. These measurements are dominated by systematic uncertainties associated with sample preparation, including purification and filtering. For extinction lengths above 10 m, the accuracy of LAB subtraction also starts to become unreliable.}
\label{Transparency}
\end{figure}

\section{Pulse Shape Discrimination}
The emission time profile of TeLS was measured using two fast 1'' square Hamamatsu PMTs, the R7600-U200 and the R5900. The R7600-U200 was used as the trigger for the DAQ and provides the time zero for the system. It was optically coupled to an UVT acrylic cube, which had a cylindrical volume hollowed out and filled with nitrogen-bubbled Te-loaded scintillator. Either a $^{90}$Sr/$^{90}$Y $\beta$ source or a $^{210}$Po $\alpha$ source was then deployed above the scintillator. The R5900 (measurement PMT) was kept 30~cm from the source to ensure a low coincidence rate, resulting in the R5900 detecting primarily single photoelectrons. A Lecroy WaveRunner 606Zi 600~MHz oscilloscope was used to digitize the signals for both PMTs. A comparison between the data for $\beta$ and $\alpha$ particles is shown in Figure 13 for scintillator samples of LAB+2g/L PPO loaded with 0.5\% Te by weight. Nearly identical distributions were obtained for either synthesis method. The difference in pulse shapes between $\beta$ and $\alpha$ particles is comparable to that seen in other liquid scintillation detectors, where good levels of PSD have been realised \cite{Borexino}.

\begin{figure}[H]
    \centering
    \includegraphics[scale=0.5]{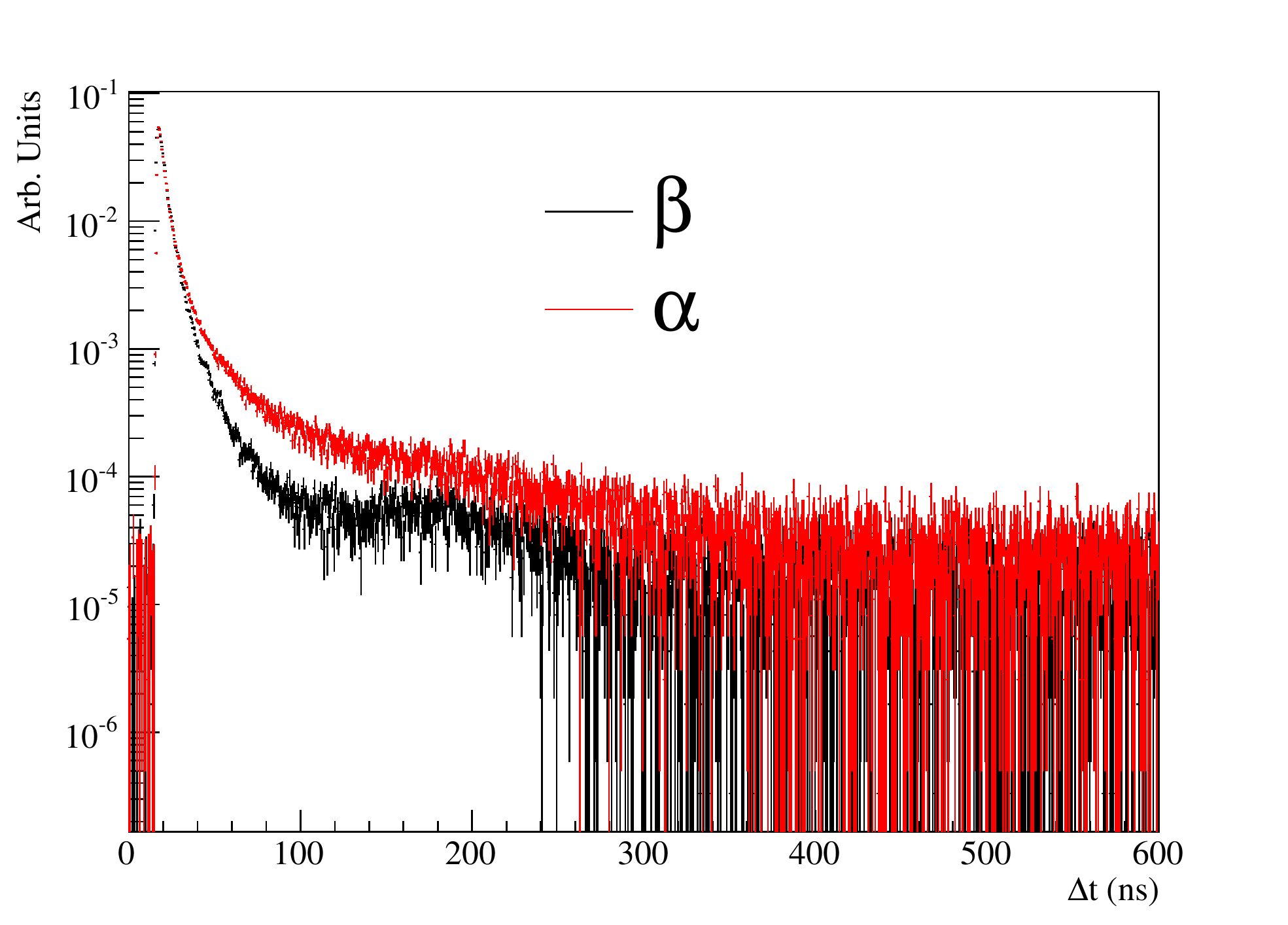}
    \caption{A comparison of the emission time profiles of 0.5\% Te-loaded LAB+PPO under $\beta$ and $\alpha$ excitation.}
    \label{fig:timing-plots}
\end{figure}

The $\Delta$t spectrum can be fit using an empirical model consisting of the sum of four exponentials:

\begin{equation}\label{eq:fit}
A_{1} \times \left[\frac{1}{\sqrt{2\pi\sigma^{2}}} e^{- \frac{ (t - \mu)^2 }{ 2\sigma^2}} \circledast \sum_{i=1}^{4}\frac{N_{i}}{\tau_{i}}e^{-t/\tau_{i}}\right] + A_{2} \times \left[\frac{1}{\sqrt{2\pi\sigma_{\rm ap}^{2}}} e^{- \frac{ (t - \mu_{\rm ap})^2 }{ 2\sigma_{\rm ap}^2}}\right] + A_{3}
\end{equation}

The first component in Equation \ref{eq:fit}, with normalization $A_{1}$, consists of a Gaussian detector response, with width $\sigma$ and mean $\mu$, convolved with the scintillator emission timing, which is modeled as the sum of four exponentials with time-constants $\tau_{i}$ and normalisations $N_{i}$, where $\sum_{i}^{4} N_{i} = 1$. The second component, with normalization $A_{2}$, models a small after-pulsing peak, with $\Delta t \approx 160$, using a Gaussian with width $\sigma_{ap}$ and mean $\mu_{ap}$. The values of $\sigma_{ap}$ and $\mu_{ap}$ are constrained based on measurements with a Cherenkov light source. The last component ($A_{3}$) models the flat dark-rate, and the values of $\sum_{i}^{3} A_{i}=1$. Fit values for the normalisations and time constants from Equation 1 for $\beta$ and $\alpha$ particle excitation are given in Table 1. 
\begin{table}[H]
\centering 
\begin{tabular}{ccccc}          
\hline\hline\noalign{\smallskip}
Particle & $N_{1}$ & $N_{2}$ & $N_{3}$ & $N_{4}$ \\
\noalign{\smallskip}\hline\noalign{\smallskip}
$\beta$ & 0.72 $\pm$ 0.02 & 0.23 $\pm$ 0.02 & 0.02 $\pm$ 0.02 & 0.03 $\pm$ 0.02 \\ \hline 
$\alpha$ & 0.63 $\pm$ 0.02 & 0.23 $\pm$ 0.02 & 0.07 $\pm$ 0.02 & 0.07 $\pm$ 0.02 \\
\noalign{\smallskip}\noalign{\smallskip}
  & $\tau_{1}$ & $\tau_{2}$ & $\tau_{3}$ & $\tau_{4}$ \\
\noalign{\smallskip}\hline\noalign{\smallskip}
$\beta$ & 3.70 $\pm$ 0.26 & 10.0 $\pm$ 2.2 & 52.0 $\pm$ 12.0 & 500 $\pm$ 176 \\ \hline 
$\alpha$ & 3.69 $\pm$ 0.10 & 15.5 $\pm$ 1.3  & 79.3 $\pm$ 10.0 & 489 $\pm$ 164 \\ \noalign{\smallskip}\hline\hline 
\end{tabular}
\caption{The fitted values for the normalisations and time constants from Equation \ref{eq:fit} for $\beta$ and $\alpha$ particle excitation of Te-loaded LAB+PPO.}
\label{tab:fit-results}
\end{table}

\section{Summary}

A  technique has been developed to efficiently load tellurium into organic liquid scintillator to provide a highly sensitive means of searching for neutrinoless double beta decay from $^{130}$Te (34\% natural abundance). This is accomplished by synthesising a group of 
compounds via condensation reactions between telluric acid and 1,2-butanediol (BD), with N,N-dimethyldodecylamine (DDA) acting as a stabilisation and solubilisation agent. Both BD and DDA can be obtained via non-biogenic production processes, which is important in order to maintain the low $^{14}$C levels needed to allow low threshold operation of large liquid scintillation detectors. The chemicals involved can be easily purified, have high flash points and can be handled underground in large quantities with reasonable safety protocols. The loading method is also found to be highly compatible with acrylic, which is important for detectors using this as a buffer between scintillation and light detection regions. Two variants of the approach have been presented: one in which solubilisation is accomplished by oligomerisation via heating, with DDA added afterwards to guarantee stability; and one in which DDA is added at the beginning to directly solubilise precursor compounds. The latter approach yields improved light output with fluorescence quenching reduced by a factor of $\sim$3 compared with the former. Both approaches result in a compound that is completely miscible with LAB scintillator with excellent optical transparency and loading stability verified to be at least in excess of 5.5 years at room temperature. The methods detailed here allow for high light output for Te loading at the level of at least several percent by weight in liquid scintillator. 

\section{Acknowledgements}
The authors wish to thank the SNO+ collaboration and SNOLAB for their support and many useful discussions. This research was supported by: {\bf Canada}: Natural Sciences and Engineering Research Council, the Canadian Institute for Advanced Research (CIFAR), Ontario Early Researcher Awards, Arthur B. McDonald Canadian Astroparticle Physics Research Institute; {\bf Germany}: the European Research Council (ERC starting grant SynPhos-307616), the German Science foundation (WE 4621/6-1) and TU Dresden; {\bf UK}: Science and Technology Facilities Council (STFC); {\bf US}: Department of Energy Office of Nuclear Physics, National Science Foundation. Appropriate representations of the data relevant to the conclusions have been provided within this paper. For the purposes of open access, the authors have applied a Creative Commons Attribution licence to any Author Accepted Manuscript version arising.

\end{document}